\documentclass[showpacs,preprintnumbers,pre,twocolumn,nofootinbib]{revtex4}
\usepackage{amsmath}
\usepackage{graphicx}
\usepackage{dcolumn}
\usepackage{bm}

\input{tcilatex}

\begin{document}

\preprint{}
\title{Noise-induced first-order transition in anti-tumor immunotherapy}
\author{Wei-Rong Zhong}
\author{Yuan-Zhi Shao}
\altaffiliation[ ]{Corresponding Author}
\email{stssyz@zsu.edu.cn}
\author{Zhen-Hui He}
\affiliation{Department of Physics, Sun Yat-sen University, 510275 Guangzhou, People's
Republic of China}

\begin{abstract}
We studied the single-variable dynamics model of the tumor growth. A
first-order phase transition induced by an additive noise is shown to
reproduce the main features of tumor growth under immune surveillance. The
critical average cells population has a power-law function relationship with
the immune coefficient.
\end{abstract}

\maketitle

During the past two decades, due to the increasing abilities of scientists
to manipulate molecule, studies of tumor immunology and immunotherapy have
entered the mainstream of current studies in immunology and cancer research
[1-3]. Simultaneously, with the development of mathematics and computation
science, more and more attempts confirm that simple math can model a world
of complexity [4]. Mathematics is then respected to bring about astonishment
into oncology. Generally, dynamics differential equations are used to
describe the growth and diffusion phenomenon of biology including tumors [5,
6]. However, there are some deficiencies for determinate differential
equation in depicting the instability and complexity of biology. Recently,
to fill these deficiencies, stochastic noise was introduced into the
dynamics equation [7, 8].

Phase transitions, especially noise-induced phase transitions, had been
reported in various areas including oncology, mathematical biology and
biological physics [7-12]. In Ref.[8] and [11], it was reported that
multiplicative noise induces a transition in tumor growth. Regretfully, two
references above have yet not specified the role of additive noise as well
as what kind of the phase transition it is. Here we explore what to our
knowledge is the first dynamical analysis of the noise-induced first-order
transition in an anti-tumor immunotherapy model.

Lefever and Garay [13] studied the tumor growth under immune surveillance
against cancer using enzyme dynamics model. The model is 
\begin{eqnarray}
&&Normal\text{ }Cells\overset{\gamma }{\rightarrow }X,  \notag \\
&&X\overset{\lambda }{\rightarrow }2X,  \notag \\
&&X+E_{0}\overset{k_{1}}{\rightarrow }E\overset{k_{2}}{\rightarrow }E_{0}+P,
\notag \\
&&P\overset{k_{3}}{\rightarrow }\text{ },  \TCItag{1}
\end{eqnarray}%
in which $X,P,E_{0}$ and $E$ are respectively cancer cells, dead cancer
cells, immune cells and the compound of cancer cells and immune cells, $%
\gamma ,\lambda ,k_{1},k_{2},k_{3}$ are velocity coefficients. This model
reveals that normal cells may transform into cancer cells, and then the
cancer cells reproduce, decline and die out ultimately. Qi and Du had ever
derived its equivalent single-variable deterministic dynamics equation [11].
Obviously, a fluctuation of tumor cells population is inevitable, which
results from the change of the environment as well as the intrinsic
instability of a tumor. Therefore, it is more reasonable to consider the
stochastic differential equation. Model (1) is then described by the
Langiven equation 
\begin{equation}
\frac{dX}{dt}=r_{0}X(1-\frac{X}{K})-\frac{\beta X}{1+X}+\xi (t)  \tag{2}
\end{equation}%
here $X$ is the cancer cells population, $r_{0}$ $,K$ and $\beta $ are the
linear per capita birth rate, the carrying capacity of the environment and
the anti-tumor ability of immune cells respectively. $\xi (t)$ is the
Gaussian white noise, $\langle \xi (t)\xi (t^{^{\prime }})\rangle =A\delta
(t-t^{^{\prime }}),$ $A$ is the intensity of noise. Since $X$ is
non-negative, according to the absorption boundary condition, then $X(t)$
can be done. We define an order parameter, $\langle X\rangle $, i.e., the
average tumor cells population, whose form is given by%
\begin{equation}
\langle X\rangle =\frac{\int_{\tau _{0}}^{\tau }X(t)dt}{\tau -\tau _{0}} 
\tag{3}
\end{equation}%
where $\tau _{0}$ is the initial time as $X(t)$ reach a stable-state
distribution. $\tau -\tau _{0}$ is the stable-state periodicity. The
"dynamic" four-order cumulant ratio $U_{X}$ is 
\begin{equation}
U_{X}=1-\frac{\langle X^{4}\rangle }{3\langle X^{2}\rangle ^{2}}  \tag{4}
\end{equation}%
which is useful for determining the location of a phase transition [14].

In a discussion of additive noise in tumor growth, Ai et al [8] reported
that additive noise leads tumor cells far from extinction. They presented
the fixedness of peak position of the stable-state distribution probability
induced by the additive noise, which denotes no transition induced by any
additive noise. Here we suggested it is because the potential function they
concerned is an uni-stable state structure. However, the equivalent
potential function of Eq.(2) is a bi-stable state one. Fig.1. shows that the
order parameter $\langle X\rangle $ is fixed under a weak noise. As the
intensity of the noise reaches an enough strength, the order parameter
undergoes a sharp transition from non-zero to zero. This will lead to a deep
minimum of the four-order cumulant ratio $U_{X}$ at the transition point. It
is obviously a first-order transition like that reported in thermodynamics.
This implies that noise could lead the tumor to extinction. The critical
point of the transition is dependent upon the immune coefficient $\beta $.%
\ifcase\msipdfoutput
\FRAME{ftbpFU}{3.3304in}{2.6446in}{0pt}{\Qcb{First-order transition of
average tumor cells population induced by noise. The parameters are $%
r_{0}=2.0,$ $K=1/3$ and $\beta =1.0$, respectively.}}{}{fig_1.eps}{%
\special{language "Scientific Word";type "GRAPHIC";maintain-aspect-ratio
TRUE;display "USEDEF";valid_file "F";width 3.3304in;height 2.6446in;depth
0pt;original-width 4.5671in;original-height 3.6192in;cropleft "0";croptop
"1";cropright "1";cropbottom "0";filename '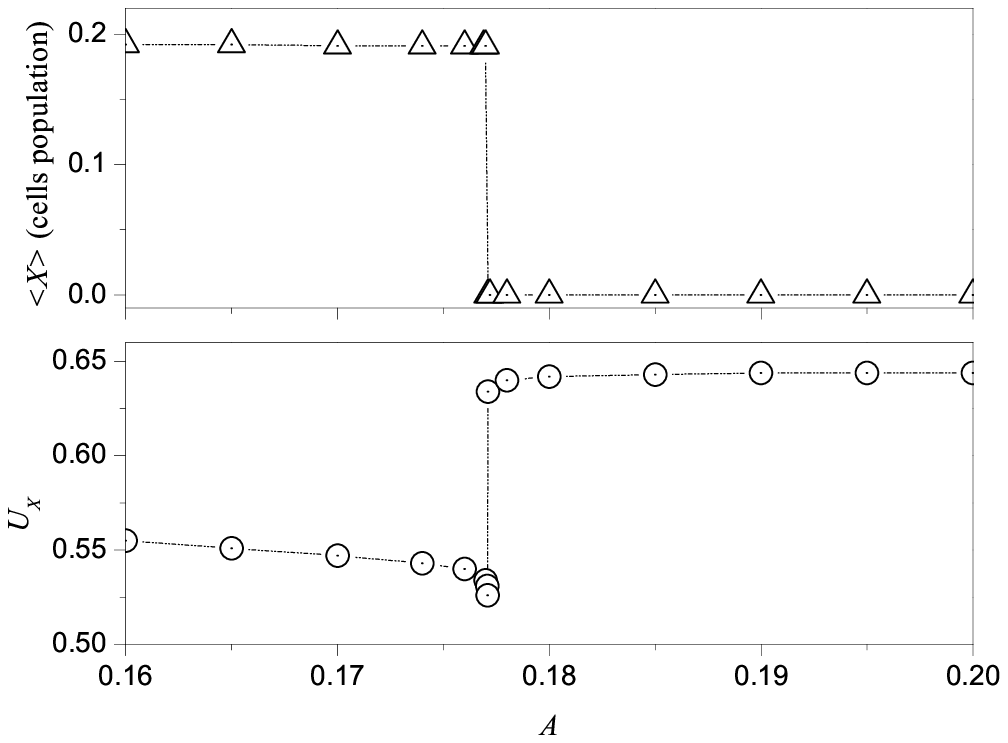';file-properties
"XNPEU";}}%
\else
\begin{figure}[ptb]\begin{center}
\includegraphics[
natheight=3.6192in, natwidth=4.5671in, height=2.6446in, width=3.3304in]
{H:/zhong/2005/Papers/NoiseImmunePRE/graphics/fig_1__1.pdf}\caption{%
First-order transition of average tumor cells population induced by noise.
The parameters are $r_{0}=2.0,$ $K=1/3$ and $\protect\beta =1.0$,
respectively.}
\end{center}\end{figure}%
\fi

In the absence of noise, $\langle X\rangle $ changes continuously with $%
\beta $ as $K<1$, i.e., there is a second-order transition induced by
parameter $\beta $ in the growth of a tumor (see Ref.[11]). In the presence
of noise, however, the variety of $\langle X\rangle $ with immune
coefficient, $\beta $, shown in Fig.2, is of a first-order transition under
an additive noise action. This is perhaps because of the noise-induced
discontinuous transition of the potential function from an uni-stable state
to a bi-stable one. The surface in Fig.2 includes two divisions: the
non-zero division of $\langle X\rangle $, here refers to an invalid
immunity; and the zero division as a valid immunity. The critical average
tumor cells population, $\langle X\rangle _{c},$ means there exists a
critical nucleus in a tumor growth, which is the minimal volume or
population for a tumor to grow and outspread.%
\ifcase\msipdfoutput
\FRAME{ftbpFU}{3.6409in}{2.6437in}{0pt}{\Qcb{Dependence of average tumor
cells population on the intensity of noise and immune coefficient.}%
}{}{fig_2.eps}{%
\special{language "Scientific Word";type "GRAPHIC";maintain-aspect-ratio
TRUE;display "USEDEF";valid_file "F";width 3.6409in;height 2.6437in;depth
0pt;original-width 8.7424in;original-height 6.3304in;cropleft "0";croptop
"1";cropright "1";cropbottom "0";filename '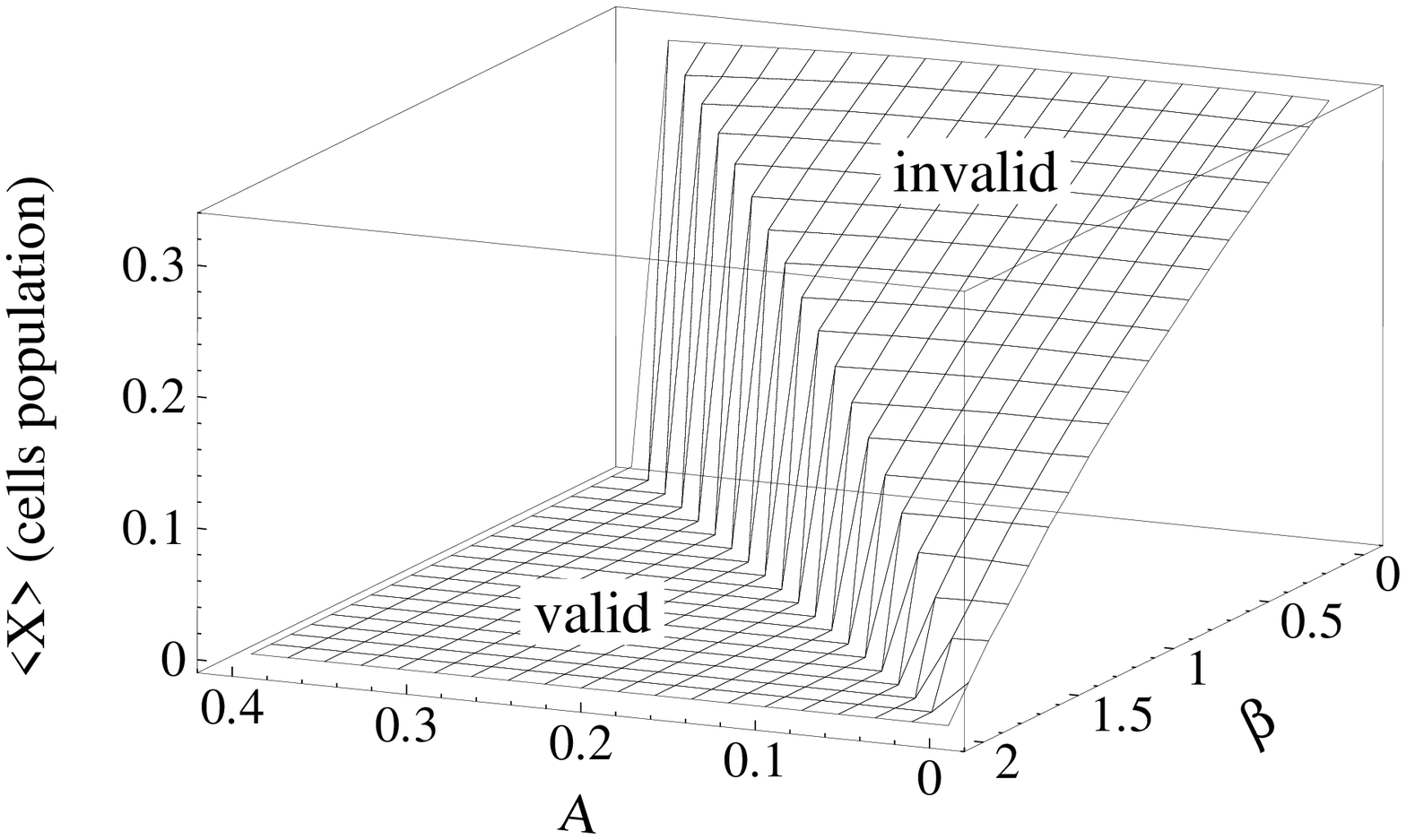';file-properties
"XNPEU";}}%
\else
\begin{figure}[ptb]\begin{center}
\includegraphics[
natheight=6.3304in, natwidth=8.7424in, height=2.6437in, width=3.6409in]
{H:/zhong/2005/Papers/NoiseImmunePRE/graphics/fig_2__2.pdf}\caption{%
Dependence of average tumor cells population on the intensity of noise and
immune coefficient.}
\end{center}\end{figure}%
\fi
\ifcase\msipdfoutput
\FRAME{ftbpFU}{3.4082in}{2.6446in}{0pt}{\Qcb{Relationship between critical
average tumor cells population and immune coefficient.}}{}{fig_3.eps}{%
\special{language "Scientific Word";type "GRAPHIC";maintain-aspect-ratio
TRUE;display "USEDEF";valid_file "F";width 3.4082in;height 2.6446in;depth
0pt;original-width 4.6363in;original-height 3.589in;cropleft "0";croptop
"1";cropright "1";cropbottom "0";filename '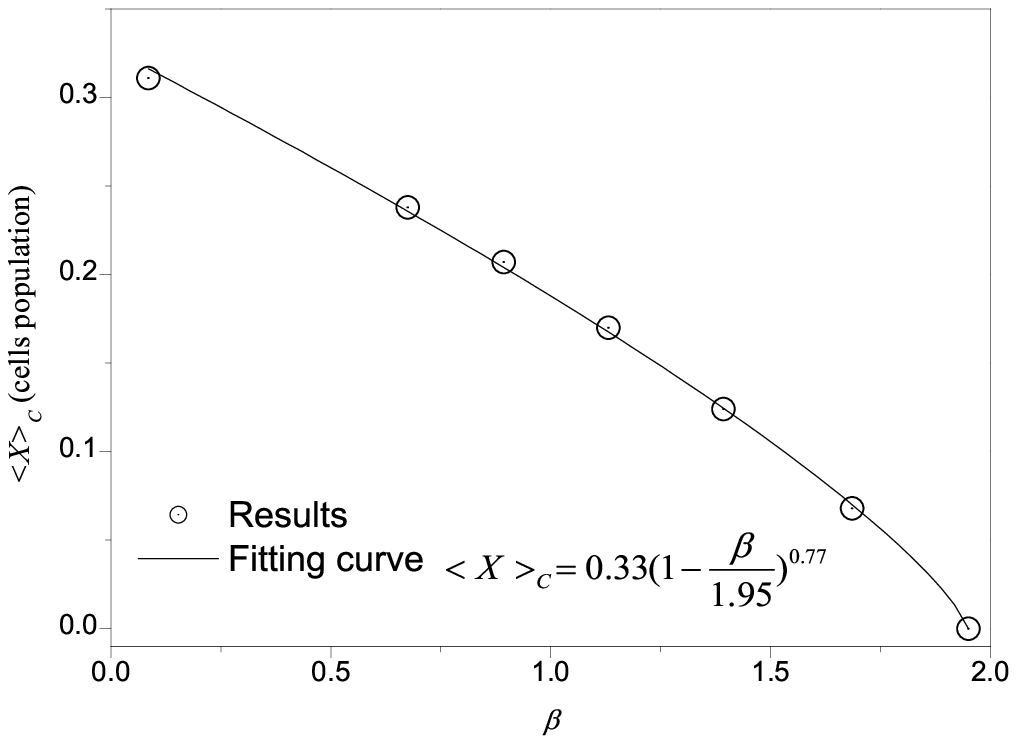';file-properties
"XNPEU";}}%
\else
\begin{figure}[ptb]\begin{center}
\includegraphics[
natheight=3.589in, natwidth=4.6363in, height=2.6446in, width=3.4082in]
{H:/zhong/2005/Papers/NoiseImmunePRE/graphics/fig_3__3.pdf}\caption{%
Relationship between critical average tumor cells population and immune
coefficient.}
\end{center}\end{figure}%
\fi

The relationship between critical average tumor cells population, $\langle
X\rangle _{c},$ and immune coefficient, $\beta $, is plotted more precisely
in Fig.3, which represents $\langle X\rangle _{c}$ as a power-law function
of $\beta ,$ which has a form with characteristic critical exponents.

\begin{equation}
\langle X\rangle _{c}=N(1-\frac{\beta }{\beta _{c}})^{\alpha }  \tag{5}
\end{equation}%
in which $N=0.33\pm 0.01,$ $\beta _{c}=1.95\pm 0.01$ and $\alpha =0.77\pm
0.03$.

In summary, additive noise will induce a first-order transition in the
growth of a tumor under immune surveillance. In other words, additive noise
can cause the decay of a tumor. There may be a critical nucleus in a tumor
growth, which is dependent on the immune coefficient as a power-law function.

This work was partially supported by the National Natural Science Foundation
(Grant No. 60471023) and the Natural Science Foundation of Guangdong
Province (Grant No. 031554), P. R. China.

\end{document}